# Tuning ultrafast electron thermalization pathways in a van der Waals heterostructure

Qiong Ma[1†], Trond I. Andersen[1†], Nityan L. Nair[1†], Nathaniel M. Gabor[1*], Mathieu Massicotte[2], Chun Hung Lui[1], Andrea F. Young[1], Wenjing Fang[3], Kenji Watanabe[4], Takashi Taniguchi[4], Jing Kong[3], Nuh Gedik[1], Frank H. L. Koppens[2], Pablo Jarillo-Herrero[1*]

†These authors contributed equally to this work.

*Correspondence to: nathaniel.gabor@ucr.edu; pjarillo@mit.edu

[1] Department of Physics, Massachusetts Institute of Technology, Cambridge, MA 02139, USA

[2] ICFO, Mediterranean Technology Park, Castelldefels (Barcelona) 08860, Spain

[3] Department of Electrical Engineering and Computer Science, Massachusetts Institute of Technology, Cambridge, MA 02139, USA

[4] National Institute for Materials Science, Namiki 1-1, Tsukuba, Ibaraki 305-0044, Japan

**Ultrafast electron thermalization - the process leading to Auger recombination[1], carrier multiplication via impact ionization[2,3], and hot carrier luminescence[4,5] - occurs when optically excited electrons in a material undergo rapid electron-electron scattering[4,6,7,8] to redistribute excess energy and reach electronic thermal equilibrium. Due to extremely short time and length scales, the measurement and manipulation of electron thermalization in nanoscale devices remains challenging even with the most advanced ultrafast laser techniques[9,10,11,12]. Here, we overcome this challenge by leveraging the atomic thinness of two-dimensional van der Waals (vdW) materials in order to introduce a highly tunable electron transfer pathway that directly competes with electron thermalization. We realize this scheme in a graphene-boron nitride-graphene (G-BN-G) vdW heterostructure[13,14,15],**

**through which optically excited carriers are transported from one graphene layer to the other. By applying an interlayer bias voltage or varying the excitation photon energy, *inter*layer carrier transport can be controlled to occur faster or slower than the *intra*layer scattering events, thus effectively tuning the electron thermalization pathways in graphene. Our findings, which demonstrate a novel means to probe and directly modulate electron energy transport in nanoscale materials, represent an important step toward designing and implementing novel optoelectronic and energy-harvesting devices with tailored microscopic properties.**

Immediately after photoexcitation of an optoelectronic device, energetic electrons scatter with other high-energy and ambient charge carriers to form a thermalized hot electron gas, which further cools by dissipating excess energy to the lattice. Due to the short distance travelled by charge carriers between electron-electron scattering events in solids[16], equilibration among the electrons occurs on the tens of femtoseconds to picosecond time scales[17, 18]. In graphene, a low-dimensional material with much enhanced Coulomb interaction[19], electron thermalization is known to occur on extremely fast time scales (<30 fs)[20, 21, 22, 23], reflecting the extremely short transit length between scattering events. Most analyses of graphene have, therefore, treated its electrons to be instantaneously thermalized[24, 25, 26, 27], and slightly non-thermal electronic behavior has thus far only been reported in pump-probe experiments with ultrashort (~10 fs) laser pulses and low excitation density[9, 10]. Due to such short time (femtosecond) and length (nanometer) scales, it is challenging to detect and control the thermalization process in graphene, or more generally, in any solid-state systems.

In this letter, we report a novel approach to probe and manipulate the electron thermalization in graphene by introducing a new energy transport channel that competes with the thermalization process. Such an additional dynamical pathway is realized in a vdW heterostructure[28] that consists of a G-BN-G stack (Figure 1a-c). In this layered structure, the photoexcited electrons in one graphene layer can travel vertically to the other graphene layer through the very thin middle BN layer (blue dashed arrow in Figure 1b). Given the close proximity of these layers, *inter*layer charge transport can occur on extremely fast time scales[29] and thus competes directly with the *intra*layer thermalization process (red arrows in Figure 1b). In our experiment, we have observed such competing processes by measuring the interlayer photocurrent under different bias and excitation conditions. Remarkably, by adjusting the interlayer bias voltage or varying the excitation photon energy, we can control the *inter*layer charge transport to occur slower or faster than the *intra*layer thermalization, thus tuning the thermalization process. Our experiments not only provide valuable insight into the electron dynamics of graphene, but also demonstrate a new means to manipulate electron thermalization in low-dimensional materials.

We fabricated the G-BN-G heterostructure devices on Si/SiO$_2$ substrates by mechanically co-laminating the graphene sheets and hexagonal boron nitride (BN) flakes[30] with 5 - 30 nm thickness (Figure 1a,d) (see Methods). In our experiment, we applied a bias voltage $V_b$ between the top and bottom graphene and measured the corresponding interlayer current $I$ under optical excitation. The main light source is a broadband supercontinuum laser that provides bright semi-continuous radiation from wavelength $\lambda$ = 450 to 2000 nm (see Methods). To probe the interlayer current in the time-domain, we also used femtosecond laser pulses from a 80-MHz Ti:Sapphire oscillator. We have measured four G-BN-G devices with monolayer graphene and two devices with few-layer (≤ four layers) graphene, and found similar results. The device characteristics are therefore insensitive to a slight change of graphene layer thickness.

Figure 1d-f show the optical image of a G-BN-G device and the corresponding device characterization using scanning photocurrent microscopy. Photocurrent images at interlayer bias voltages $V_b$ = -0.5 and 0.5 V under 600-nm laser excitation show that the photocurrent $I$ appears only in the area where the two graphene layers overlap with each other. The photocurrent magnitude increases significantly with $V_b$, and its direction flips with the sign of bias voltage (Figure 1g). We do not observe any current in the absence of illumination, indicating that optical excitation is necessary to generate a measurable interlayer current in our G-BN-G devices. Based on recent first-principle calculations[31], the graphene Dirac point is located at $\Delta$ ~ 1.3 eV above the BN valence band edge, and ~ 4.5 eV below the conduction band edge (shown schematically in Figure 1c), in agreement with dark tunneling measurements[13, 14]. This suggests that, since the potential energy barrier $\Delta$ is much smaller for positive charge carriers (holes) than for electrons, the interlayer current is mediated predominantly by holes.

The photocurrent in our G-BN-G devices exhibits complex dependence on excitation laser power $P$, interlayer bias $V_b$, and excitation photon energy $\hbar\omega$ (Figure 2). While the photocurrent generally increases with laser power, it exhibits both linear and superlinear power dependence, depending sensitively on bias voltage and photon energy. If we keep a constant bias $V_b$ = 5 V, $I$ increases superlinearly with $P$ at $\hbar\omega$ = 1.75 eV (red dots in the left panel of Figure 2a), yet gradually becomes linear as the photon energy increases to $\hbar\omega$ = 2.43 eV (purple squares). If we instead keep a constant photon energy $\hbar\omega$ = 2.10 eV, the photocurrent increases superlinearly with P at $V_b$ = 1.5 V (red dots in the right panel of Figure 2a), then gradually becomes linear as the bias increases to $V_b$ = 10 V (purple squares).

Remarkably, the complexity of the photocurrent variations can be efficiently captured by a single fitting parameter $\gamma$, which we obtained by fitting all photocurrent vs. power data with a simple power law $I \sim P^\gamma$. Figure 2b, our main result, shows $\gamma$ over a wide range of bias voltage and photon energy. The data is separated into several distinct regions, labeled A and B, highlighting the superlinear ($\gamma > 1$, red-yellow color) and linear ($\gamma = 1 \pm 0.02$, blue) power dependence of the photocurrent, respectively. The value of γ demonstrates a clear and gradual transition from roughly three at low bias and photon energy (region A) to γ ~ 1 at high bias voltage and photon energy (region B) (see Supplementary Information)

We attribute the complex photocurrent behavior to the transition between two distinct processes, thermionic emission and direct carrier tunneling, both of which mediate charge carrier transit through the G-BN-G heterostructure (Figure 2c). In thermionic emission, the photo-excited carriers remain in the graphene, scatter with one another, and quickly reach thermal equilibrium among themselves. High-energy carriers in the hot tails of the resultant thermal distribution have sufficient energy to overcome the potential energy barrier $\Delta$, and will travel to the other graphene layer[32, 33, 34]. While previous studies[27, 35] have shown that the temperature of hot carriers in graphene scales with laser power approximately as $T \sim P^{1/3}$, the population of thermionically emitted carriers increases exponentially with the temperature for the high BN barrier. We therefore expect an overall superlinear dependence of the photocurrent on the laser power[32] (see Supplementary Information). This behavior matches well with our observation at low $V_b/\hbar\omega$ (Regime A in Figure 2b-c).

In contrast, at high $V_b/\hbar\omega$, the effective BN barrier is reduced, allowing the excited carriers to tunnel from one graphene layer to the other before they scatter with other carriers. Given the energy height and thickness of the BN potential energy barrier, optical excitation is necessary to assist the tunneling process[13, 14]. Photon-assisted tunneling current then scales with the number of initial photo-excited carriers and therefore increases linearly with laser power. This behavior matches well with the observed linear power dependence of photocurrent at high $V_b/\hbar\omega$ (Regime B in Figure 2b-c).

To confirm that electron thermalization dominates the photocurrent response in the thermionic regime, we performed photocurrent measurements using a Ti:Sapphire femtosecond laser at low $V_b/\hbar\omega$. At a photon energy of 1.55 eV, photo-excitation with short pulses (90-fs duration) produces photocurrent that is orders of magnitude higher than that with a supercontinuum laser (pulse duration ~90 ps) at the same fluence, indicating that shorter pulses produce higher transient electronic temperature. Additionally, we measured the photocurrent under photo-excitation by two identical laser pulses with orthogonal polarization (Methods) and varying temporal separation[27, 35]. We observed strong positive two-pulse correlation in the photocurrent signal, which exhibits a short component (<100 fs) and a long component (~1 ps) (Figure 3 and Methods). These results, consistent with thermionic emission, are similar to the previously reported two-pulse correlation of hot photoluminescence[4], a phenomenon that arises from hot carriers at the high-energy tail of the thermal distribution in graphene. Moreover, positive correlation in the two-pulse photocurrent measurement immediately excludes direct carrier tunneling with a linear *P*-dependence as the major photocurrent mechanism at low $V_b$. We thus conclude that the observed photocurrent at low $V_b/\hbar\omega$ arises from high-energy thermalized carriers in graphene.

In the photon-assisted tunneling regime (high bias voltage and photon energy), electron tunneling is described by the Fowler-Nordheim (FN) formalism. For electrons tunneling through a triangular barrier in the presence of a high electric field, the current-voltage characteristics take the form[36]:

$$I(V) \propto V_b^2 \exp\left[-\frac{\beta}{V_b}\right], \qquad (1)$$

$$\text{where } \beta = -\frac{4d\sqrt{2m}(\Delta - \hbar\omega/2)^{3/2}}{3e\hbar}. \qquad (2)$$

Here *m* is the carrier effective mass in the barrier region, and the width *d* is determined by the BN thickness (see Supplementary Information).

The FN model exhibits several features that serve as fingerprints of a non-thermal carrier tunneling process, and can be compared directly with experiment. Equation (1) suggests a linear relationship between ln*(I/V²_b)* and *1/V*_b, with a bias-independent slope $\beta$. To examine the regime over which this behavior holds, we analyzed the *I-V*_b data taken under supercontinuum laser excitation. Figure 4a shows ln*(I/V²_b)* vs. *1/V*_b at positive bias over a range of photon energies. At high bias or photon energy (Region B in Figure 2c), we observed a linear relationship that breaks down at low $V_b/\hbar\omega$, where thermionic emission dominates (Region A). We extracted the slope (-$\beta$) from the linear fits, and found that $\beta^{2/3}$ scales linearly with the excitation photon energy $\hbar\omega$, consistent with Equation (2) (red dots in Figure 4b). From fitting the data and comparing to Equation (2), we estimate the barrier height from the *x*-axis intercept to be Δ = 1.25 eV. Similar results were obtained at negative bias (see Supplementary Information), with a slightly higher estimated barrier height Δ = 1.31 eV (blue dots in Figure 4b). These values agree well with the predicted potential barrier (~1.3 eV) between graphene and BN by first-principle calculations[31], and confirm that the photocurrent is dominated by direct carrier tunneling at high $V_b/\hbar\omega$.

By adjusting the interlayer bias voltage and excitation photon energy, we can tune the interlayer transport process, thus allowing us to pinpoint the regime in which both processes occur on similar time scales. For direct carrier tunneling, the average carrier transit time $\tau_{tun}$ can be estimated as $\tau_{tun} = \tau/T(E) \sim h/ET(E)$, where $\tau \sim h/E$ is approximated from the uncertainty principle (see Supplementary Information), $h$ is Planck's constant, $E$ is the energy difference between initial and final states, and $T(E)$ is the WKB transmission probability[36, 37]. Both $T(E)$ and $\tau$ depend on the excitation energy and bias voltage, while $T(E)$ also contains information about the barrier height and effective mass in BN (see Supplementary Information). As a function of bias voltage and photon energy, this model predicts that the tunneling time remains constant over a series of lines in $\hbar\omega$ vs. $V_b$ space. Figure 2b shows several of these lines (black dashed lines) and the corresponding average carrier transit times ($\tau_{tun}$ = 2, 7, 100 and 1000 fs). Carrier tunneling occurs faster at high $V_b/\hbar\omega$ than at low $V_b/\hbar\omega$, with tunneling times ranging from of $\tau_{tun}$ = 1 fs ~ almost infinity (near zero bias voltage). Our estimate of the transit times exhibits quite remarkable agreement with experiment. In particular, the contour at $\tau_{tun}$ = 7 fs effectively captures the main transition (blue to red) between features in our photocurrent image. Furthermore, the fit implies a carrier thermalization time on the order of 10 fs in graphene, which indeed matches excellently the results from other ultrafast experiments[21].

In the regime for which the thermalization time matches the tunneling time (white-red areas in Figure 2b), we also observed peculiar behaviors in the $I$-$V_b$ characteristics, indicating a strong transition between direct tunneling and thermionic emission. Particularly, we observed an abrupt change of linearity between $\ln(I/V_b^2)$ and $1/V_b$ (Figure 4a), which signifies the breaking down of the FN approximation as $V_b/\hbar\omega$ decreases and hence a diminishing contribution from non-thermal carrier tunneling. Intriguingly, at the onset of FN tunneling, weak oscillations in the $\ln(I/V_b^2)$ vs $1/V_b$ plot can be seen. These can be more easily observed by plotting $d^2I/dV_b^2$, which changes between negative and positive values as $V_b/\hbar\omega$ increases (blue and red stripes in Figure 4c). We also observe an additional replica oscillation feature that appears parallel to, yet ~200 meV above, the major oscillation feature in the $d^2I/dV_b^2$ map. These subtle features may indicate resonance effects in the FN tunneling regime, whose origin could be related to field emission resonances[38] due to spatial confinement or resonant phonon emission[36, 39, 40]. More work, beyond the scope of this manuscript, will be needed to investigate in depth the origin of these resonances.

In conclusion, we have demonstrated that interlayer photocurrent in a G-BN-G heterostructure arises from two competing ultrafast processes, thermionic emission and direct carrier tunneling, which dominate the photocurrent at low and high interlayer bias/photon energy, respectively. The interplay between these two channels allows us to tune direct carrier tunneling so that it occurs faster or slower than the *intra*layer electron-electron scattering process in graphene, thus tuning the electron thermalization pathways. With appropriate modeling, we have deduced the thermalization time in graphene (~10 fs) from a time-averaged transport experiment. Similar experimental methods can readily be extended to other van der Waals heterostructures to probe their ultrafast electronic processes. More generally, given the central role that electron thermalization plays in energy transport, the tunability on femtosecond time scales demonstrated in our research opens up a new range of in-situ functionality for novel optoelectronic devices.

## Methods

**Device fabrication.** We fabricated G-BN-G heterostructure devices on $Si/SiO_2$ substrates. The graphene layers were prepared either by mechanical exfoliation of graphite or by chemical vapor deposition (CVD) on copper surface. In the heterostructure, the bottom layer was either exfoliated graphene, or transferred CVD graphene that were patterned into strips by e-beam lithography and $O_2$ plasma etching. The middle BN flake was exfoliated from high-quality bulk crystals onto Methyl methacrylate (MMA) polymer and transferred onto the bottom graphene. The top graphene layer was similarly exfoliated onto MMA polymer and transferred onto the BN. Finally we deposited 0.8/80-nm Cr/Au electrodes with e-beam lithography and thermal evaporation.

**Photocurrent measurements.** We have carried out the photocurrent experiment in a confocal microscope. The devices are mounted in an optical cryostat cooled by liquid helium, with controllable temperature in the range of 4-300 K. The excitation beam comes from a broadband supercontinuum fiber laser (Fianium). The beam, with tunable wavelength through a monochromator, is focused onto the samples with a spot diameter of ~1 μm. By using a piezoelectrically controlled mirror, we can scan the beam across the whole device area to obtain photocurrent images.

**Two-pulse correlation measurements.** We have carried out time-domain photocurrent measurements with a 80-MHz Ti:Sapphire oscillator (Tsunami) that generates femtosecond laser pulses with central wavelength 800 nm. The laser is separated into two equally intense beams with controllable path-length difference, and focused onto the devices with a spot diameter of ~2 μm. Photocurrent was measured as a function of temporal separation between the two pulses. To suppress the interference at zero time delay, the polarizations of the two beams were set to be orthogonal to each other by a half-wave plate. A set of neutral density filters was used to adjust the pump fluence. We have characterized the pulse duration with second harmonic autocorrelation. From the autocorrelation width (~130 fs), we determined a pulse duration of ~90 fs at our sample position. The response time of the devices is extracted by fitting the photocurrent correlation data with a symmetric biexponential function convoluted with a Gaussian function of width ~130 fs.

**Acknowledgements**

We thank V. Fatemi, L. Ju, L. Levitov, J. Rodriguez-Nieva, J. Sanchez-Yamagishi, E. J. Sie, J. C. W. Song and H. Steinberg for discussions. This work was supported by AFOSR Grant No. FA9550-11-1-0225 (measurement and data analysis, Q.M., T.A., N.N., N.G., and P.J.H.) and the Packard Fellowship program. This work made use of the Materials Research Science and Engineering Center Shared Experimental Facilities supported by the National Science Foundation (NSF) (Grant No. DMR-0819762) and of Harvard's Center for Nanoscale Systems, supported by the NSF (Grant No. ECS-0335765). N.G. and C.H.L. have been supported by the Gordon and Betty Moore Foundation's EPiQS Initiative through Grant GBMF4540 for the time-domain photocurrent measurements. F.K. acknowledges support by the Fundacio Cellex Barcelona, the ERC starting grant 307806 (CarbonLight), and support by the EE Graphene

Flagship (contract No. 604391). W.F. and J.K. acknowledge the funding support by the STC Center for Integrated Quantum Materials, NSF Grant No. DMR-1231319.

**Author Contributions**

NMG and PJH conceived the experiment. QM, NLN and MM fabricated the devices. NLN, NMG, QM and MM carried out the spatial and spectral photocurrent measurements. TIA and CHL performed the time-domain photocurrent measurements under the supervision from NG. QM, TIA, NLN and MM analyzed the data under supervision from NMG, CHL, AFY, FHLK, and PJH. WF and JK grew the CVD graphene. KW and TT synthesized the BN crystals. QM, TIA, CHL, NLN, NMG, FHKL, and PJH cowrote the paper with input from all other authors.

**Author Information**

Reprints and permissions information is available at…. The authors declare no competing financial interests. Readers are welcome to comment on the online version of the paper. Correspondence and requests for materials should be addressed to P.J.H.

Figures and Captions

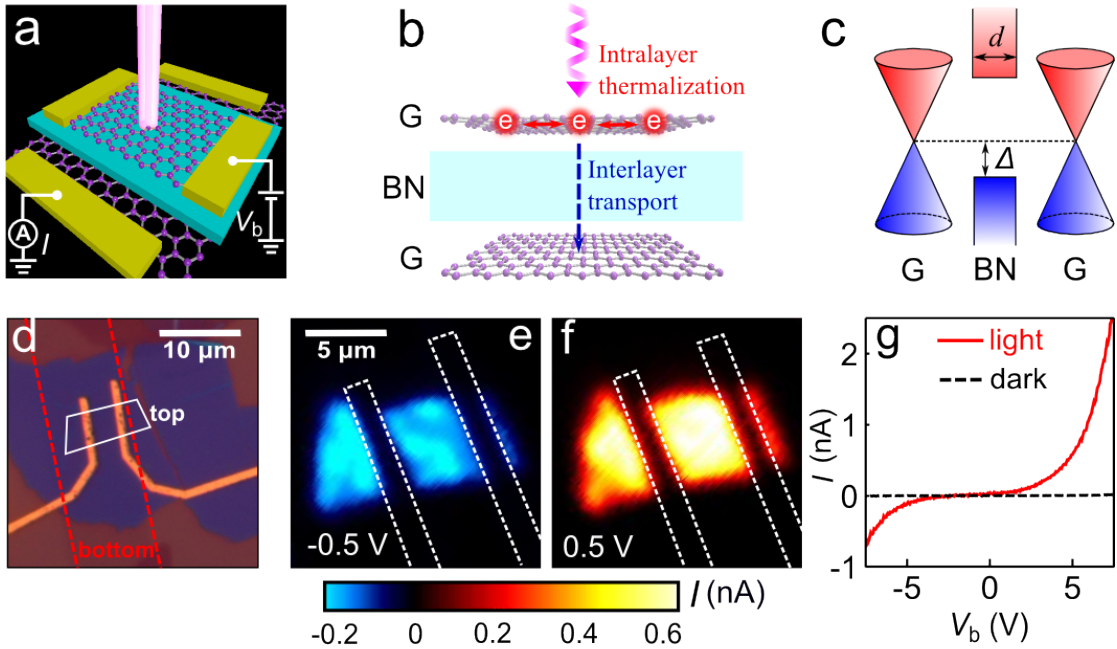

**Figure 1| Interlayer photocurrent of a G-BN-G device. a,** Schematic of a G-BN-G device under optical excitation. **b,** Schematic of *intra*layer thermalization and *inter*layer transport of the optically excited carriers. **c,** Band alignment between graphene and BN. BN has a band gap of ~5.9 eV, and the Dirac point of graphene is located ~1.3 eV ($\Delta$) above the edge of the BN valence band[31] . **d**, Optical image of a G-BN-G device, which consists of a top exfoliated graphene layer (white line), a 14-nm-thick BN flake, and a bottom graphene layer grown by chemical vapor deposition (dashed red line). **e-f**, Scanning images of interlayer photocurrent at interlayer bias $V_b$ = -0.5 and 0.5 V. **g**, Interlayer photocurrent as a function of $V_b$ with and without light illumination. All measurements were carried out with 600-nm optical excitation from a supercontinuum laser at $T$ = 100 K. The incident laser power is 500 µW for (e-f) and 100 µW for (g).

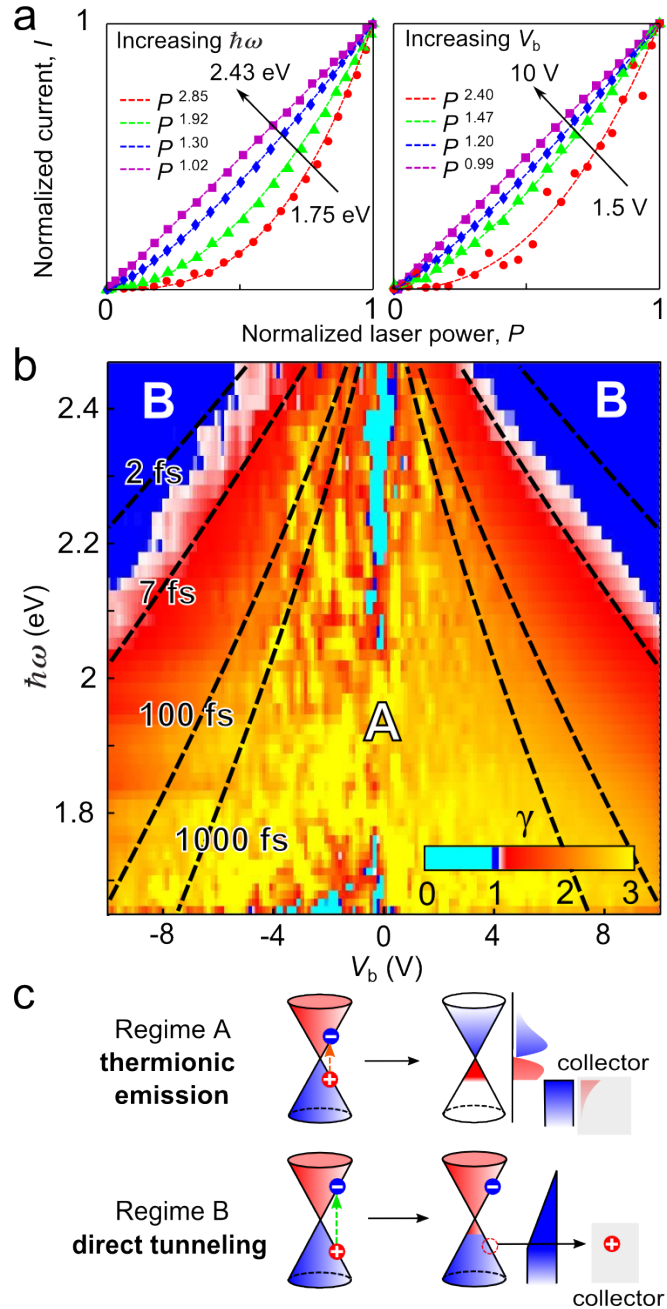

**Figure 2| Two different regimes of interlayer photocurrent in a G-BN-G device. a,** Photocurrent *I* as a function of excitation laser power *P*, at constant interlayer bias $V_b = 5$ V but increasing photon energies $\hbar\omega$ = 1.75, 2.10, 2.25 and 2.43 eV (left panel), and at constant $\hbar\omega$ = 2.10 eV but increasing bias $V_b$ = 1.5, 3.75, 5 and 10 V (right panel). For better comparison, the current and laser power are normalized, and the data are fitted with a power law $I \sim P^\gamma$. The raw data are shown in the Supplementary Information. **b,** A color map of $\gamma$ as a function of $V_b$ and $\hbar\omega$. The color scale is customized to make the area with $\gamma > 1$ (Region A) and $\gamma = 1 \pm 0.02$ (Region B) appear red-yellow and blue, respectively. The black dashed lines are contours corresponding to tunneling time of 2, 7, 100 and 1000 fs, predicted by our model described in the text. **c,** Schematics depicting the thermionic emission and the direct carrier tunneling as the

dominant photocurrent mechanism for Region A and B in (c), respectively. All measurements were carried out with a supercontinuum laser at $T = 100$ K.

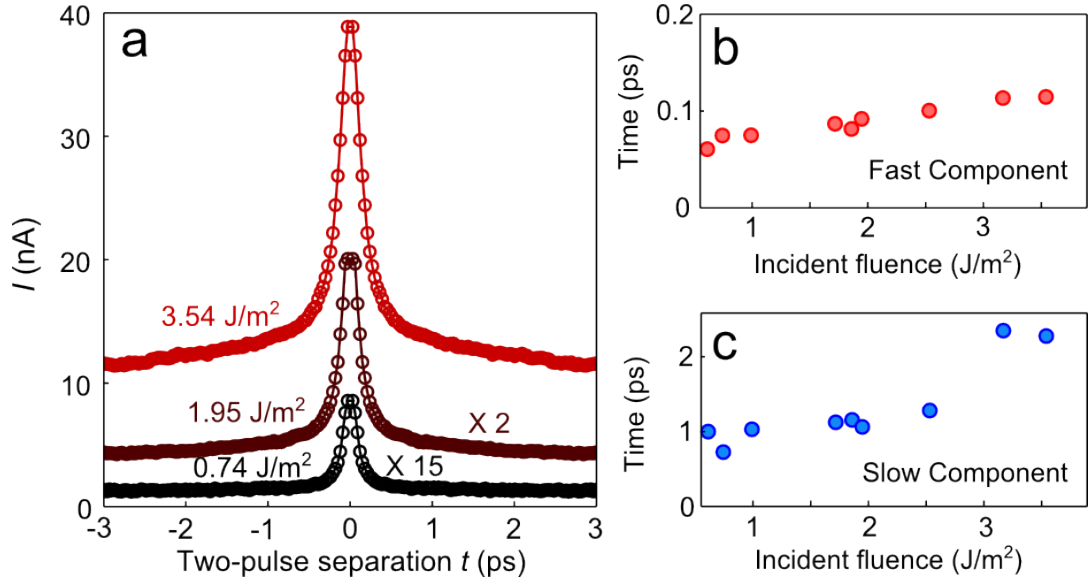

**Figure 3| Two-pulse correlation of interlayer photocurrent in a G-BN-G device. a,** Photocurrent at $V_b = 0.1$ V, as a function of temporal separation between two identical but cross-polarized excitation pulses. The incident fluences of each beam are 0.74, 1.95 and 3.54 J/cm$^2$. The pulse duration is 90 fs and the photon energy is 1.55 eV. **b-c,** The time constants of the fast and slow components at different incident fluences. They are extracted by fitting the correlation data with a symmetric biexponential function convoluted with a Gaussian function of width 130 fs. All measurements were carried out with a Ti:Sapphire laser at $T = 300$ K.

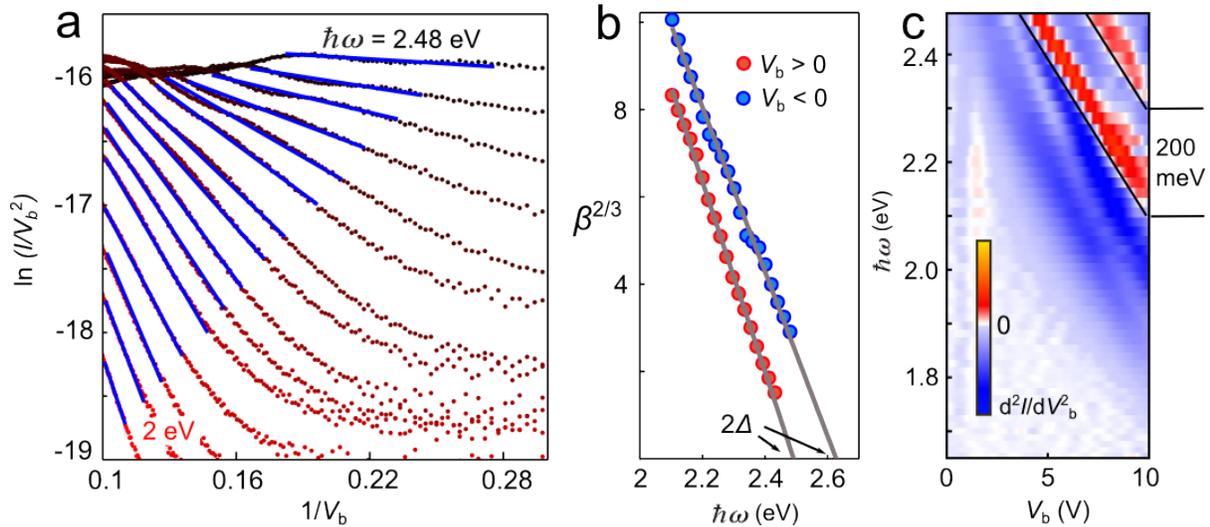

**Figure 4| Experimental signatures of direct Fowler-Nordheim carrier tunneling. a**, $\ln(I/V^2_b)$ as a function of $1/V_b$ at different excitation photon energy $\hbar\omega$ = 2.1 to 2.45 eV. The blue lines highlight the linear behavior at high bias, with slope of $\beta$. **b**, $\beta^{2/3}$ as a function of $\hbar\omega$ at positive bias (red dots) and negative bias (blue dots). The lines are linear fits with *x*-intercepts at $2\Delta$, where $\Delta$ = 1.25 and 1.31 eV correspond to the barrier height in the Fowler-Nordheim formula, eq.(2) in the text. **c**, $d^2I/dV^2_b$ map as a function of $V_b$ and $\hbar\omega$. The dashed lines highlight two oscillation features with a separation of 200 meV. All measurements were carried out with a supercontinuum laser at $T$ = 100 K.